\documentclass[fleqn]{LxSymp_2022_Paper_Format}
\usepackage{amsmath,bm}
\usepackage{tikz}
\usepackage{placeins}
\usepackage{subfigure}
\newcommand\figref{Fig.~\ref}
\begin{document}

\title{A RANS approach to the Meshless Computation of Pressure Fields From\\ Image Velocimetry}
\firstpagehead{}
\runningheads{20th LISBON Laser Symposium 2022}{}
\runnigfoots{}

\author{Pietro Sperotto$^{1*}$,  Sandra Pieraccini$^2$,  Miguel A. Mendez$^1$}

\address{	$^1$von Karman Institute for Fluid Dynamics, Sint-Genesius-Rode, Belgium \\%
	$^2$Dipartimento di Ingegneria Meccanica e Aerospaziale, Politecnico di Torino, Italy\\
\vspace{7pt}
\vskip 1pt
{\addressize{*Corresponding author:
\href{mailto:pietro.sperotto@vki.ac.be}{\textstyleInternetlink{pietro.sperotto@vki.ac.be}}}}}
\fontfamily{ppl}\selectfont
\linespread{1.2}
\vskip 3 mm
\begin{center}
\end{center}
\vskip 8 mm
\abstract{
We propose a 3D meshless method to compute mean pressure fields in turbulent flows from image velocimetry. The method is an extension of the constrained Radial Basis Function (RBF) formulation by \citet{Sperotto2022} to a Reynolds Averaged Navier Stokes (RANS) framework. This is designed to handle both scattered data as in Particle Tracking Velocimetry (PTV) and data in uniform grids as in correlation-based Particle Image Velocimetry (PIV). The RANS extension includes the Reynolds stresses into the constrained least square problem. We test the approach on a numerical database featuring a Backward Facing Step (BFS) with a Reynolds number of 6400 (defined with respect to the inlet velocity and step height), obtained via Direct Numerical Simulation (DNS).
}

\section{Introduction}\label{Sec:Intro}

Many methods have been proposed to compute pressure fields from image velocimetry in the last decade. These can be broadly classified into Eulerian and Lagrangian methods depending on how the material acceleration is computed. 
Eulerian methods are based on the available velocity fields, which can be instantaneous or time-averaged. These can be further classified into directional method, integrating Navier Stokes equation (eg. \cite{Jakobsen1997,Kngeter1999PIVWH,Liu2006,Wang2019}), and global method, solving the pressure Poisson equation (eg. \cite{gurka1999computation,Ghaemi2012,Pan2016}). Lagrangian methods are based on the computation of particle trajectories and have been primarily fostered by recent advances in 3D tracking techniques (\cite{Schanz2016}). These can be further distinguished into techniques that interpolate the particle acceleration onto a Cartesian grid \citep{Geseman,Huhn2016,Huhn2018,Schneiders2016} and techniques that integrate the pressure on scattered data such as the Voronoi integration proposed by \cite{Neeteson2015}.

All of the aforementioned methods rely on ‘classic‘ mesh-based numerical methods (e.g. Finite Differences or Finite Elements) to solve the Partial Differential Equations (PDE) involved, namely the Navier Stokes Equation or the Poisson equation. Recently, we proposed a meshless approach based on Radial Basis Functions \citep{Sperotto2022} for integrating the pressure field from scattered velocity measurements in an incompressible and laminar flow. This approach builds on the vast literature of RBF based PDE solvers \citep{Kansa1990,Fornberg2015,cmes,Chen2002,Chen2003,Sarler} and consists in solving two linear least square problems with differential constraints. The first is the regression of the velocity field constrained by the continuity equation and the relevant boundary conditions; the second is the integration of the pressure Poisson equation given the regressed velocity fields.

In this work, we extend the previous contribution in \cite{Sperotto2022} to the Reynolds Averaged Navier Stokes (RANS) formulation firstly proposed by \citet{gurka1999computation}. The pressure integration consists in solving the pressure Poisson equation in the RANS problem, i.e. including the contribution of the Reynolds stresses. Assuming that the available dataset is statistically sufficiently resolved to compute the Reynolds stresses, these are regressed using an RBF expansion as done for the velocity fields. Both the RBF approximations of the Reynolds stresses and velocity fields are then used in the pressure computation.

We introduce the RANS based pressure computation problem and the RBF-based integration approach in Section \ref{Sec2}. We test our approach on the flow past a Backward Facing Step (BFS) at a moderately large Reynolds number. The dataset was constructed by sampling a Direct Numerical Simulation (DNS), thus providing high-fidelity and statistically resolved data on the mean velocity and mean pressure fields. This allows testing the accuracy of the RBF regression and pressure computation. The test case is described in Section 3. Finally, section 4 reports on the comparison between the RBF regression and pressure integration and the ground truth, while section 5 closes with conclusions and outlooks.

\section{Methodology}\label{Sec2}

The starting point of the method is the RANS formulation, which we briefly recall in section \ref{RANS}. Three linear least square problems are set. The first two are (1) the RBF regression of the (mean) velocity field and (2) the RBF regression of the Reynolds stresses. Both are carried out with differential constraints. The third problem consists in (3) solving the pressure Poisson equation using the (analytic) results of the previous two steps. 

While the solution of problem (1) and (3) are extensively described in \cite{Sperotto2022}, the step (2) constitute a major novelty and is briefly illustrated in section \ref{Reynolds}. The fundamentals of the RBF-regression and meshless Poisson solver are briefly recalled in \ref{RBF}.

\subsection{The RANS formulation}\label{RANS}

The RANS equation are obtained by introducing the Reynolds decomposition $\boldsymbol{u}=\overline{\boldsymbol{u}}(\boldsymbol{x})+\boldsymbol{u}'(\boldsymbol{x},t)$, where $\overline{\boldsymbol{u}}(\mathbf{x})=(\overline{u},\overline{v},\overline{w})$ is the averaged velocity field and $\boldsymbol{u}'(\boldsymbol{x},t)=(u',v',w')$ is the fluctuating (time dependent) field. For the incompressible flow of a Newtonian fluids with constant properties, introducing this decomposition in the Navier stokes equation and averaging yields the RANS equations:
\begin{ceqn}
\begin{equation}
\label{eq:RANS}
    \rho\left(\overline{\boldsymbol{u}}\cdot \nabla\right)\overline{\boldsymbol{u}}=-\nabla \overline{p} + \mu \Delta\overline{\boldsymbol{u}}-\rho \nabla \cdot \mathbf{R}\left(\boldsymbol{u}',\boldsymbol{u}'\right),
\end{equation}\end{ceqn} where $p$ the pressure field, $\rho$ is the density, $\mu$ the dynamic viscosity and $\mathbf{R}\left(\boldsymbol{u}',\boldsymbol{u}'\right)=\overline{u'_iu'_j}$ is the Reynolds stresses tensor, with the indices $i,j=1,2,3$ referring to the three components. This collects the correlation between the fluctuating components of the velocity field and introduces additional stresses used to model the mean flow's non-linear diffusion due to turbulence. Taking the divergence on both sides and using the continuity equation for the mean flow (i.e. $\nabla\cdot\overline{\boldsymbol{u} }=0$) yields the pressure Poisson equation in the RANS context. Together with its boundary conditions, this reads \cite{Pope2000}
\begin{ceqn}
\begin{equation}
\label{eq:PoissonA}
    \begin{cases}
        \Delta \overline{p}=-\rho \left(\nabla \cdot \left(\overline{\boldsymbol{u}}\cdot\nabla \overline{\boldsymbol{u}}\right)+\nabla\cdot\left(\nabla\cdot \mathbf{R}\left(\boldsymbol{u}',\boldsymbol{u}'\right)\right)\right) \quad \boldsymbol{x}\in \Omega\\
        \partial_{\boldsymbol{n}} \overline{p}=\nabla \overline{p} \cdot \boldsymbol{n}=\rho \left(-\left(\overline{\boldsymbol{u}} \cdot \nabla\right) \overline{\boldsymbol{u}}-\nabla \cdot \mathbf{R}\left(\boldsymbol{u}',\boldsymbol{u}'\right)+\nu \Delta \overline{\boldsymbol{u}}\right)\cdot \boldsymbol{n}\quad \boldsymbol{x}\in \Gamma_N\\
        \overline{p}=\overline{p_D} \quad \boldsymbol{x} \in \Gamma_D.
    \end{cases}
\end{equation}\end{ceqn} where $\Omega \subset \mathbb{R}^3$ is the integration domain with boundaries $\partial \Omega$. The Dirichlet and Neumann condition are applied respectively in $\Gamma_D$ and $\Gamma_N$, which are such that $\Gamma_D\cup\Gamma_N=\partial \Omega$.

In the context of pressure integration from measured velocity fields, the Dirichlet conditions are usually prescribed by a set of pressure probes (e.g. pressure taps or Pitot tubes) while the Neumann boundary conditions are evaluated by projecting the RANS equations into the relevant boundaries. The reader is referred to \cite{Pan2016,Faiella2021} for an extensive discussion on the impact of boundary conditions in the pressure integration from image velocimetry.

\subsection{RBF Regression and Meshless PDE Integration}\label{RBF}

The RBF regression consists in finding an approximation of a function as a linear combination of RBFs. That is, given a function ${f}(\boldsymbol{x}): \mathbb{R}^3\rightarrow \mathbb{R}$, we seek the set of weights $w_{i}$ such that  

\begin{equation}
 \label{eq:labelapprox}
   {f}(\boldsymbol{x})\simeq \tilde{f}(\boldsymbol{x})=\sum_{i=1}^{n_{c}} w_{i} \varphi\left(\boldsymbol{x} \mid \boldsymbol{x_i}^{*}, c_{i}\right)=\sum_{i=1}^{n_{c}} w_{i} \varphi_{i} 
\end{equation} where $c_i$ and $\boldsymbol{x_i}^*$ are the shape factor and the collocation point of the i-th basis functions $\varphi_i$ and $n_c$ is the number of bases involved. As in \cite{Sperotto2022}, we here consider isotropic Gaussians as basis functions.

Assuming that values of $f(\boldsymbol{x})$ are given in a set of $n_p$ points such that $\boldsymbol{f}=f\left(\boldsymbol{x}_j\right)=f_j$ with $j=1,\dots,n_p$, the regression consists in finding the weights $w_{i}$ that minimize the least square error between the approximation and the available points. Arranging the basis functions into a matrix $\mathbf{\Phi}\left(\boldsymbol{X}\right)\in \mathbb{R}^{n_p\times n_c}$, with $\boldsymbol{X}\in \mathbb{R}^{3\times n_p}$ the coordinates of the points $\mathbf{x}_j$ in which data is available, equation \eqref{eq:labelapprox} can be written as $\boldsymbol{f}\approx \mathbf{\Phi}\left(\boldsymbol{X}\right)\cdot\boldsymbol{w}$ where $\boldsymbol{w}$ is the vector collecting the $n_c$ weights $w_i$.

Denoting as $\mathcal{L} \{f\}(\mathbf{x})$ a differential operator acting on $f$, one has $\mathcal{L} \{f\}(\mathbf{x})=\mathcal{L} \{\mathbf{\Phi}\}\left(\boldsymbol{x}\right) \cdot\boldsymbol{w}$. A differential constraint is thus a linear constraint on $\boldsymbol{w}$, here denoted $\mathcal{L} \{\mathbf{\Phi}\} \cdot\boldsymbol{w}=\boldsymbol{f}_L$. Similarly, Dirichlet boundary conditions on $\mathbf{X}_D\in\Gamma_D$ can be written as $\mathbf{\Phi}\left(\mathbf{X}_D\right)\cdot \boldsymbol{w}=\boldsymbol{f}_D$ and Neumann boundary conditions on $\mathbf{X}_N\in\Gamma_N$ can be written as $\partial_n\mathbf{\Phi}\left(\mathbf{X}_N\right)\cdot \boldsymbol{w}=\boldsymbol{f}_N$, where $\partial_n$ denotes derivatives normal to the boundaries. 


The RBF regression with differential constraints with Dirichlet and Neumann boundary conditions is thus a quadratic programming problem with linear constraints. Using Lagrange multipliers, the solution consists in finding the weights $\boldsymbol{w}$ and the multipliers $\boldsymbol{\lambda}=(\boldsymbol{\lambda}_{\mathcal{L}},\boldsymbol{\lambda}_{D},\boldsymbol{\lambda}_{N})$ that minimize the augmented cost function

\begin{equation}
 \label{eq:G}
    J^*(\boldsymbol{w},\bm{\lambda})=||\boldsymbol{f}-\boldsymbol{\Phi} (\boldsymbol{X})\cdot \boldsymbol{w}||^2_2+\bm{\lambda}^T_{\mathcal{L}}\biggl(\mathcal{L} \{\mathbf{\Phi}\} \cdot\boldsymbol{w}-\boldsymbol{f}_L\biggr)+
    \bm{\lambda}^T_{\mathcal{D}}\biggl(\mathbf{\Phi}\left(\Gamma_D\right)\cdot \boldsymbol{w}-\boldsymbol{f}_D\biggr)+
    \bm{\lambda}^T_{\mathcal{N}}\biggl(\mathbf{\Phi}\left(\Gamma_N\right)\cdot \boldsymbol{w}-\boldsymbol{f}_N\biggr)
\end{equation} where $||\bullet||_2$ denotes the $l_2$ norm.

In the regression of the velocity field, differential constraints are used to impose the divergence-free conditions; the Dirichlet conditions are used to set no-slip on walls and Neumann conditions are used on 'outlet' patches. In the solution of the pressure Poisson equation, the basis functions are the Laplacian of the RBFs, and the constraints are used to set the boundary conditions for the pressure integration. The reader is referred to \cite{Sperotto2022} for more details on the least square problem formulation for scalar and vector fields and the numerical methods implemented for their solution. 

\subsection{The Regression of the Reynolds stresses}\label{Reynolds}

The problem of regressing the Reynolds stresses is analogous to the problem of regressing the velocity field. In the formulation implemented in this work, no simplifications are considered for the Reynolds stresses. Therefore, this regression consists in regressing six functions, that is the entries in the tensor $\mathbf{R}\left(\boldsymbol{u}',\boldsymbol{u}'\right)=\overline{u'_iu'_j}$.
The problem can be considerably simplified if turbulence is assumed to be isotropic (then $\mathbf{R}$ is diagonal) or isotropic \emph{and} homogeneous (then $\mathbf{R}$ is a multiple the identity matrix). 

The regression of these terms were constrained with the Dirichlet condition $\overline{u'_iu'_j}=0 \,\forall i,j=1,2,3$ at walls. Moreover, since the velocity fluctuations must vanish sufficiently close to the walls, it is also possible to show that all derivatives of the Reynolds stresses must also vanish at walls, that is $\partial_{{x}_k} \overline{u'_iu'_j} =0 \,\forall i,j,k=1,2,3$.

\section{Select Test Case}\label{Sec3}

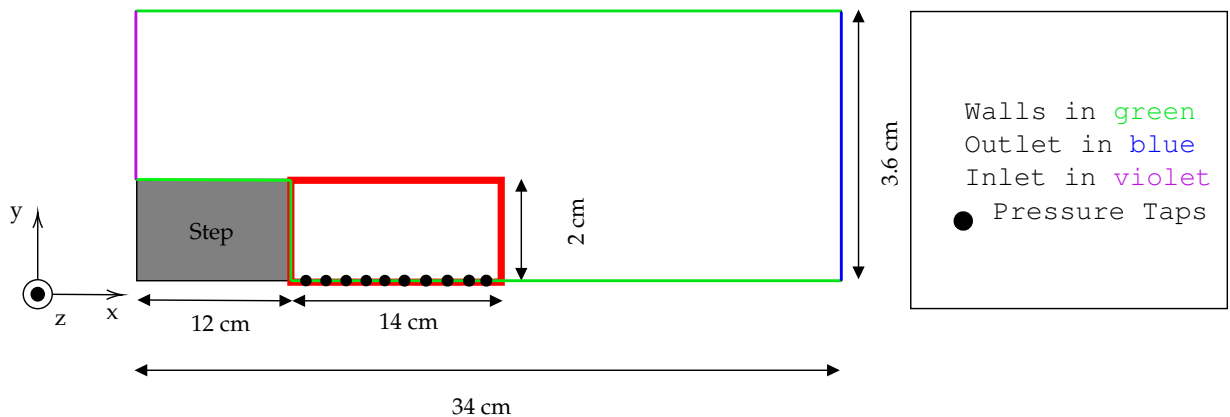
\begin{figure}[h!]
\centering

\tikzset{every picture/.style={line width=0.75pt}} 
\resizebox{0.9\textwidth}{!}{

\tikzset{every picture/.style={line width=0.75pt}} 

\tikzset{every picture/.style={line width=0.75pt}} 

\begin{tikzpicture}[x=0.75pt,y=0.75pt,yscale=-1,xscale=1]

\draw   (100,41.6) -- (591.6,41.6) -- (591.6,159) -- (100,159) -- cycle ;
\draw   (208,159.6) -- (591.6,159.6) -- (591.6,229.6) -- (208,229.6) -- cycle ;
\draw [color={rgb, 255:red, 255; green, 255; blue, 255 }  ,draw opacity=1 ][line width=5.25]    (208,159.6) -- (591.6,159) ;
\draw    (212,243.6) -- (352.6,243.6) ;
\draw [shift={(355.6,243.6)}, rotate = 180] [fill={rgb, 255:red, 0; green, 0; blue, 0 }  ][line width=0.08]  [draw opacity=0] (8.93,-4.29) -- (0,0) -- (8.93,4.29) -- cycle    ;
\draw [shift={(209,243.6)}, rotate = 0] [fill={rgb, 255:red, 0; green, 0; blue, 0 }  ][line width=0.08]  [draw opacity=0] (8.93,-4.29) -- (0,0) -- (8.93,4.29) -- cycle    ;
\draw    (102,292) -- (587.6,292) ;
\draw [shift={(590.6,292)}, rotate = 180] [fill={rgb, 255:red, 0; green, 0; blue, 0 }  ][line width=0.08]  [draw opacity=0] (8.93,-4.29) -- (0,0) -- (8.93,4.29) -- cycle    ;
\draw [shift={(99,292)}, rotate = 0] [fill={rgb, 255:red, 0; green, 0; blue, 0 }  ][line width=0.08]  [draw opacity=0] (8.93,-4.29) -- (0,0) -- (8.93,4.29) -- cycle    ;
\draw    (104,243.02) -- (206,243.58) ;
\draw [shift={(209,243.6)}, rotate = 180.32] [fill={rgb, 255:red, 0; green, 0; blue, 0 }  ][line width=0.08]  [draw opacity=0] (8.93,-4.29) -- (0,0) -- (8.93,4.29) -- cycle    ;
\draw [shift={(101,243)}, rotate = 0.32] [fill={rgb, 255:red, 0; green, 0; blue, 0 }  ][line width=0.08]  [draw opacity=0] (8.93,-4.29) -- (0,0) -- (8.93,4.29) -- cycle    ;
\draw    (603.6,43.6) -- (603.6,225.6) ;
\draw [shift={(603.6,228.6)}, rotate = 270] [fill={rgb, 255:red, 0; green, 0; blue, 0 }  ][line width=0.08]  [draw opacity=0] (8.93,-4.29) -- (0,0) -- (8.93,4.29) -- cycle    ;
\draw [shift={(603.6,40.6)}, rotate = 90] [fill={rgb, 255:red, 0; green, 0; blue, 0 }  ][line width=0.08]  [draw opacity=0] (8.93,-4.29) -- (0,0) -- (8.93,4.29) -- cycle    ;
\draw    (368.6,161.6) -- (368.6,226.6) ;
\draw [shift={(368.6,229.6)}, rotate = 270] [fill={rgb, 255:red, 0; green, 0; blue, 0 }  ][line width=0.08]  [draw opacity=0] (8.93,-4.29) -- (0,0) -- (8.93,4.29) -- cycle    ;
\draw [shift={(368.6,158.6)}, rotate = 90] [fill={rgb, 255:red, 0; green, 0; blue, 0 }  ][line width=0.08]  [draw opacity=0] (8.93,-4.29) -- (0,0) -- (8.93,4.29) -- cycle    ;
\draw  [fill={rgb, 255:red, 128; green, 128; blue, 128 }  ,fill opacity=1 ] (100.6,159.6) -- (208,159.6) -- (208,229.6) -- (100.6,229.6) -- cycle ;
\draw  [color={rgb, 255:red, 255; green, 0; blue, 0 }  ,draw opacity=1 ][line width=3.75]  (208,159.6) -- (354.6,159.6) -- (354.6,230.6) -- (208,230.6) -- cycle ;
\draw    (31.6,239) -- (31.6,183.8) ;
\draw [shift={(31.6,181.8)}, rotate = 90] [color={rgb, 255:red, 0; green, 0; blue, 0 }  ][line width=0.75]    (10.93,-3.29) .. controls (6.95,-1.4) and (3.31,-0.3) .. (0,0) .. controls (3.31,0.3) and (6.95,1.4) .. (10.93,3.29)   ;
\draw    (31.6,239) -- (88.6,239.39) ;
\draw [shift={(90.6,239.4)}, rotate = 180.39] [color={rgb, 255:red, 0; green, 0; blue, 0 }  ][line width=0.75]    (10.93,-3.29) .. controls (6.95,-1.4) and (3.31,-0.3) .. (0,0) .. controls (3.31,0.3) and (6.95,1.4) .. (10.93,3.29)   ;
\draw  [fill={rgb, 255:red, 255; green, 255; blue, 255 }  ,fill opacity=1 ] (21.5,239) .. controls (21.5,233.42) and (26.02,228.9) .. (31.6,228.9) .. controls (37.18,228.9) and (41.7,233.42) .. (41.7,239) .. controls (41.7,244.58) and (37.18,249.1) .. (31.6,249.1) .. controls (26.02,249.1) and (21.5,244.58) .. (21.5,239) -- cycle ;
\draw  [fill={rgb, 255:red, 0; green, 0; blue, 0 }  ,fill opacity=1 ] (27.06,239) .. controls (27.06,236.42) and (29.09,234.34) .. (31.6,234.34) .. controls (34.11,234.34) and (36.14,236.42) .. (36.14,239) .. controls (36.14,241.58) and (34.11,243.66) .. (31.6,243.66) .. controls (29.09,243.66) and (27.06,241.58) .. (27.06,239) -- cycle ;
\draw [color={rgb, 255:red, 0; green, 251; blue, 11 }  ,draw opacity=1 ][line width=1.5]    (100,159) -- (208,159.6) ;
\draw [color={rgb, 255:red, 0; green, 226; blue, 34 }  ,draw opacity=1 ][line width=1.5]    (208,159.6) -- (208,229.6) ;
\draw [color={rgb, 255:red, 0; green, 226; blue, 18 }  ,draw opacity=1 ][line width=1.5]    (208,229.6) -- (591.6,229.6) ;
\draw [color={rgb, 255:red, 0; green, 226; blue, 26 }  ,draw opacity=1 ][line width=1.5]    (100,41.6) -- (591.6,41.6) ;
\draw [color={rgb, 255:red, 189; green, 16; blue, 224 }  ,draw opacity=1 ][line width=1.5]    (100,159) -- (100,41.6) ;
\draw [color={rgb, 255:red, 0; green, 0; blue, 255 }  ,draw opacity=1 ][line width=1.5]    (591.6,229.6) -- (591.6,42.2) ;
\draw   (640,42) -- (860.6,42) -- (860.6,249) -- (640,249) -- cycle ;
\draw  [fill={rgb, 255:red, 0; green, 0; blue, 0 }  ,fill opacity=1 ] (215,229.5) .. controls (215,227.57) and (216.57,226) .. (218.5,226) .. controls (220.43,226) and (222,227.57) .. (222,229.5) .. controls (222,231.43) and (220.43,233) .. (218.5,233) .. controls (216.57,233) and (215,231.43) .. (215,229.5) -- cycle ;
\draw  [fill={rgb, 255:red, 0; green, 0; blue, 0 }  ,fill opacity=1 ] (229,229.5) .. controls (229,227.57) and (230.57,226) .. (232.5,226) .. controls (234.43,226) and (236,227.57) .. (236,229.5) .. controls (236,231.43) and (234.43,233) .. (232.5,233) .. controls (230.57,233) and (229,231.43) .. (229,229.5) -- cycle ;
\draw  [fill={rgb, 255:red, 0; green, 0; blue, 0 }  ,fill opacity=1 ] (243,229.5) .. controls (243,227.57) and (244.57,226) .. (246.5,226) .. controls (248.43,226) and (250,227.57) .. (250,229.5) .. controls (250,231.43) and (248.43,233) .. (246.5,233) .. controls (244.57,233) and (243,231.43) .. (243,229.5) -- cycle ;
\draw  [fill={rgb, 255:red, 0; green, 0; blue, 0 }  ,fill opacity=1 ] (257,229.5) .. controls (257,227.57) and (258.57,226) .. (260.5,226) .. controls (262.43,226) and (264,227.57) .. (264,229.5) .. controls (264,231.43) and (262.43,233) .. (260.5,233) .. controls (258.57,233) and (257,231.43) .. (257,229.5) -- cycle ;
\draw  [fill={rgb, 255:red, 0; green, 0; blue, 0 }  ,fill opacity=1 ] (270,229.5) .. controls (270,227.57) and (271.57,226) .. (273.5,226) .. controls (275.43,226) and (277,227.57) .. (277,229.5) .. controls (277,231.43) and (275.43,233) .. (273.5,233) .. controls (271.57,233) and (270,231.43) .. (270,229.5) -- cycle ;
\draw  [fill={rgb, 255:red, 0; green, 0; blue, 0 }  ,fill opacity=1 ] (288.58,232.94) .. controls (286.7,233.71) and (284.56,232.81) .. (283.79,230.93) .. controls (283.03,229.04) and (283.93,226.9) .. (285.81,226.14) .. controls (287.69,225.37) and (289.83,226.27) .. (290.6,228.15) .. controls (291.36,230.04) and (290.46,232.18) .. (288.58,232.94) -- cycle ;
\draw  [fill={rgb, 255:red, 0; green, 0; blue, 0 }  ,fill opacity=1 ] (303.58,232.94) .. controls (301.7,233.71) and (299.56,232.81) .. (298.79,230.93) .. controls (298.03,229.04) and (298.93,226.9) .. (300.81,226.14) .. controls (302.69,225.37) and (304.83,226.27) .. (305.6,228.15) .. controls (306.36,230.04) and (305.46,232.18) .. (303.58,232.94) -- cycle ;
\draw  [fill={rgb, 255:red, 0; green, 0; blue, 0 }  ,fill opacity=1 ] (318.58,232.94) .. controls (316.7,233.71) and (314.56,232.81) .. (313.79,230.93) .. controls (313.03,229.04) and (313.93,226.9) .. (315.81,226.14) .. controls (317.69,225.37) and (319.83,226.27) .. (320.6,228.15) .. controls (321.36,230.04) and (320.46,232.18) .. (318.58,232.94) -- cycle ;
\draw  [fill={rgb, 255:red, 0; green, 0; blue, 0 }  ,fill opacity=1 ] (333.58,232.94) .. controls (331.7,233.71) and (329.56,232.81) .. (328.79,230.93) .. controls (328.03,229.04) and (328.93,226.9) .. (330.81,226.14) .. controls (332.69,225.37) and (334.83,226.27) .. (335.6,228.15) .. controls (336.36,230.04) and (335.46,232.18) .. (333.58,232.94) -- cycle ;
\draw  [fill={rgb, 255:red, 0; green, 0; blue, 0 }  ,fill opacity=1 ] (345.58,232.94) .. controls (343.7,233.71) and (341.56,232.81) .. (340.79,230.93) .. controls (340.03,229.04) and (340.93,226.9) .. (342.81,226.14) .. controls (344.69,225.37) and (346.83,226.27) .. (347.6,228.15) .. controls (348.36,230.04) and (347.46,232.18) .. (345.58,232.94) -- cycle ;
\draw  [fill={rgb, 255:red, 0; green, 0; blue, 0 }  ,fill opacity=1 ] (678.95,193.79) .. controls (675.74,195.09) and (672.08,193.55) .. (670.77,190.34) .. controls (669.46,187.12) and (671.01,183.46) .. (674.22,182.15) .. controls (677.43,180.85) and (681.1,182.39) .. (682.4,185.6) .. controls (683.71,188.82) and (682.17,192.48) .. (678.95,193.79) -- cycle ;

\draw (319,310) node [anchor=north west][inner sep=0.75pt]   [align=left] {34 cm};
\draw (268,251) node [anchor=north west][inner sep=0.75pt]   [align=left] {14 cm};
\draw (399.62,206.56) node [anchor=north west][inner sep=0.75pt]  [rotate=-269.59] [align=left] {2 cm};
\draw (137,252) node [anchor=north west][inner sep=0.75pt]   [align=left] {12 cm};
\draw (618.1,165.04) node [anchor=north west][inner sep=0.75pt]  [rotate=-269.75] [align=left] {3.6 cm};
\draw (136,188) node [anchor=north west][inner sep=0.75pt]   [align=left] {Step};
\draw (78,246.8) node [anchor=north west][inner sep=0.75pt]   [align=left] {x};
\draw (11,176.8) node [anchor=north west][inner sep=0.75pt]   [align=left] {y};
\draw (42,251.8) node [anchor=north west][inner sep=0.75pt]   [align=left] {z};
\draw (676,103) node [anchor=north west][inner sep=0.75pt]   [align=left] {{\fontfamily{pcr}\selectfont {\large Walls in \textcolor[rgb]{0,0.89,0.1}{green}}}\\{\fontfamily{pcr}\selectfont {\large Outlet in \textcolor[rgb]{0,0,1}{blue}}}\\{\fontfamily{pcr}\selectfont {\large Inlet in \textcolor[rgb]{0.74,0.06,0.88}{violet}}}\\ \ \ \ \ \ {\large {\fontfamily{pcr}\selectfont Pressure Taps}}\\ \ \ \ };

\end{tikzpicture}

}
\caption{Domain of the analyzed configuration, namely the flow past a BFS via DNS by \citet{ODER2019118436}. The region of interest simulating an experimental acquisition is shown in red. The black markers illustrate the position of $10$ pressure taps, installed with a spacing of $1.4$cm, used to aid the pressure computation from the velocity fields.}
\label{fig:comp}
\end{figure}

We create a synthetic 3D PTV database for the flow past a Backward Facing Step (BFS). The data was obtained via Direct Numerical Simulation (DNS) by \citet{ODER2019118436}. This simulation analyzes the heat transfer in a low Prandtl number fluid ($Pr=0.005$) and Reynolds number, based on the inlet velocity $U_0$ and step height $h$, of $Re=\rho h U_0/\mu=6400$. The main geometrical parameters are shown in Figure \ref{fig:comp}.

The dataset was provided in dimensionless form, with velocities normalized with respect to the inlet velocity and spatial coordinates normalized with respect to the step height. Therefore, we take the freedom of setting the inlet velocity to $U_0=1$m/s, and the step height to $h=2$cm. Taking $\rho=1$ kg/m$^3$, we then consider $\mu/\rho=3.125\cdot10^{-5}$m$^2$/s to keep the same Reynolds number. The remaining relevant dimensions, as shown in Figure \ref{fig:comp}, respect the proportions of the computational domain by \citet{ODER2019118436}. The domain extends $3.6$ cm in the $z$ direction. Inlet, outlet and wall patches are shown with violet, blue and green lines respectively. 

We construct the 3D PTV dataset assuming that the region of interest for the measurement is the one in red in figure \ref{fig:comp}. The measurement volume is $8$mm thick in the z-direction and is located in the centre of the channel along z. To aid the pressure computation from the image velocimetry, we assume that $10$ pressure taps are installed on the bottom wall (black markers in figure \ref{fig:comp})  with a spacing of $1.4$cm.

\begin{figure}[h!]
\centering
\includegraphics[clip,trim=0 0.25cm 0cm 0,width=15cm]{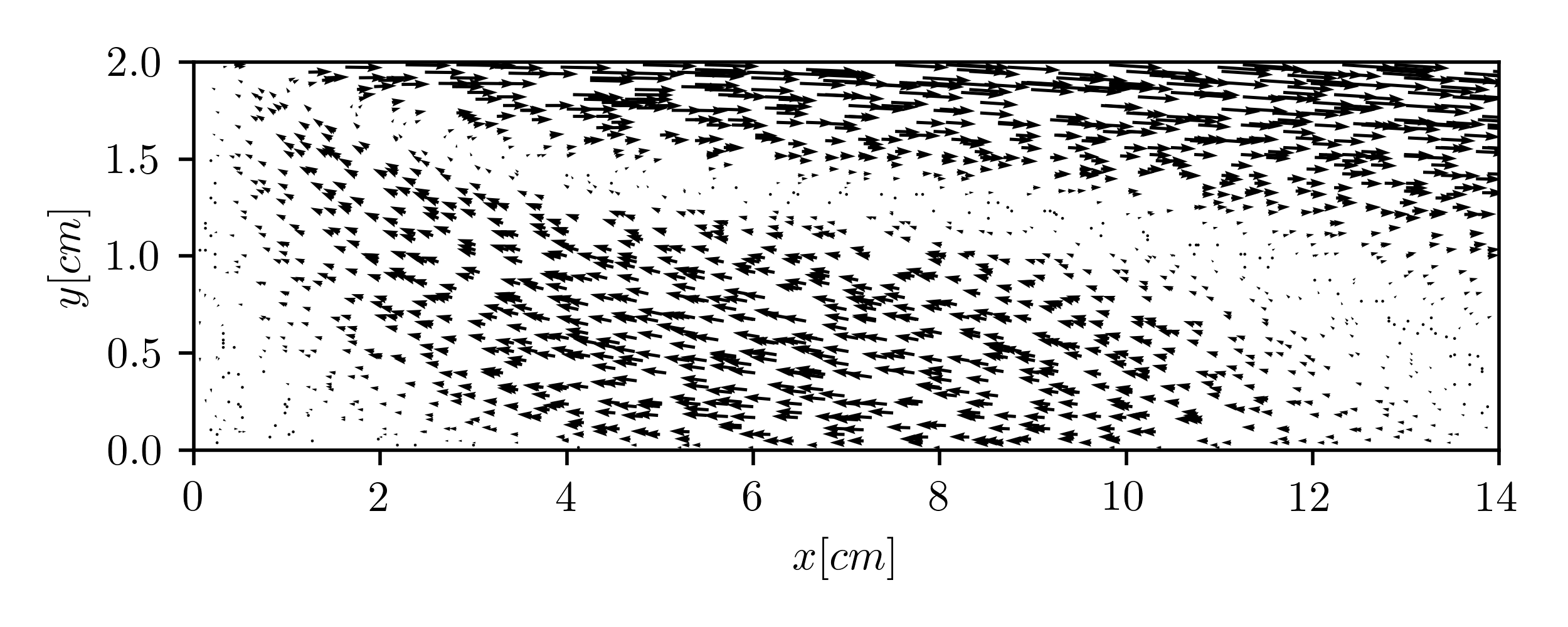}
\caption{Plot of the sampled mean flow field in the middle plane. All the available points are shown; this allows visualizing the (modest) seeding density used in the investigated test case. }
\label{quiver}
\end{figure}

This work considers only averaged quantities, hence mean velocity, mean pressure, and Reynolds's stress fields. These were randomly sampled on $8400$ points from the available computational grid, i.e. avoiding any form of interpolation in the data generation. Considering images with 1024 x 2048 pixels, this leads to a particle density of $3.4\cdot10^{-3}$ ppp. Figure \ref{quiver} shows the mean velocity field of the obtained PTV dataset. This scattered sampling of the averaged fields can be pictured as the result of an ensemble PTV averaging as proposed by \citet{aguera2016ensemble}. Our meshless method works equally well on a uniform grid (such as obtained via PIV), but we stick to the scattered sampling approach to test the accuracy of the regression without the interfering effect of interpolation in the data preparation. Similarly, we here take the Reynolds stresses in these points from the DNS to evaluate the regression performances on the ground truth data. However, to make the data more realistic, we add a random white noise of $5$\% on the velocity, pressure and Reynolds stresses fields as well as on the sampling points simulating pressure taps.

\section{Results}

The results are given in terms of the euclidean norm $E=\|f\left(\mathbf{X}\right)-\tilde{f}\left(\mathbf{X}\right)\|_2$ and the relative euclidean norm $\varepsilon=\|f\left(\mathbf{X}\right)-\tilde{f}\left(\mathbf{X}\right)\|_2/\|f\left(\mathbf{X}\right)\|_2$ in Table \ref{tab:Errors}. These result show that some variables ($\overline{w},\overline{u'w'}$ and $\overline{v'w'}$) have large values of $\varepsilon$. However, these fields still an acceptable $E$, hence the main discrepancy is due to the low value of these quantities for the investigated test case. 

\begin{figure}
    \centering
    \includegraphics[width=0.48\textwidth]{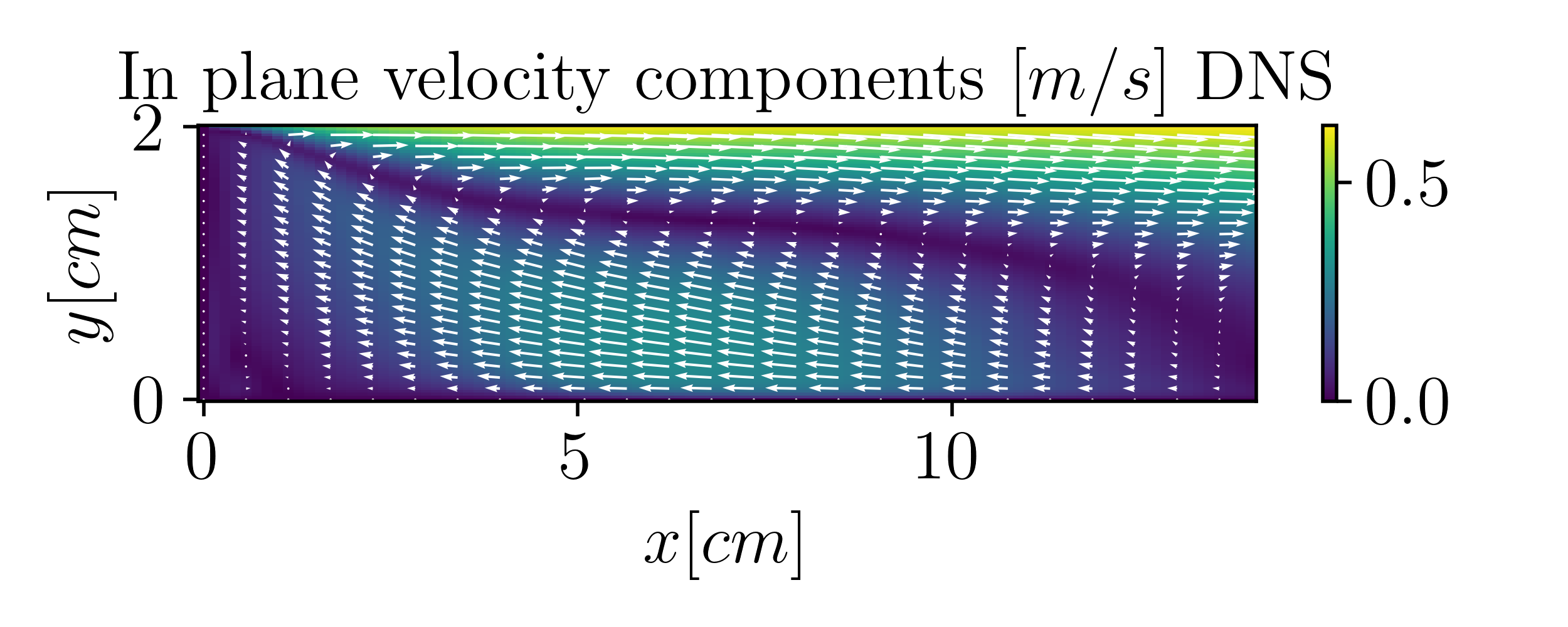}
    \includegraphics[width=0.48\textwidth]{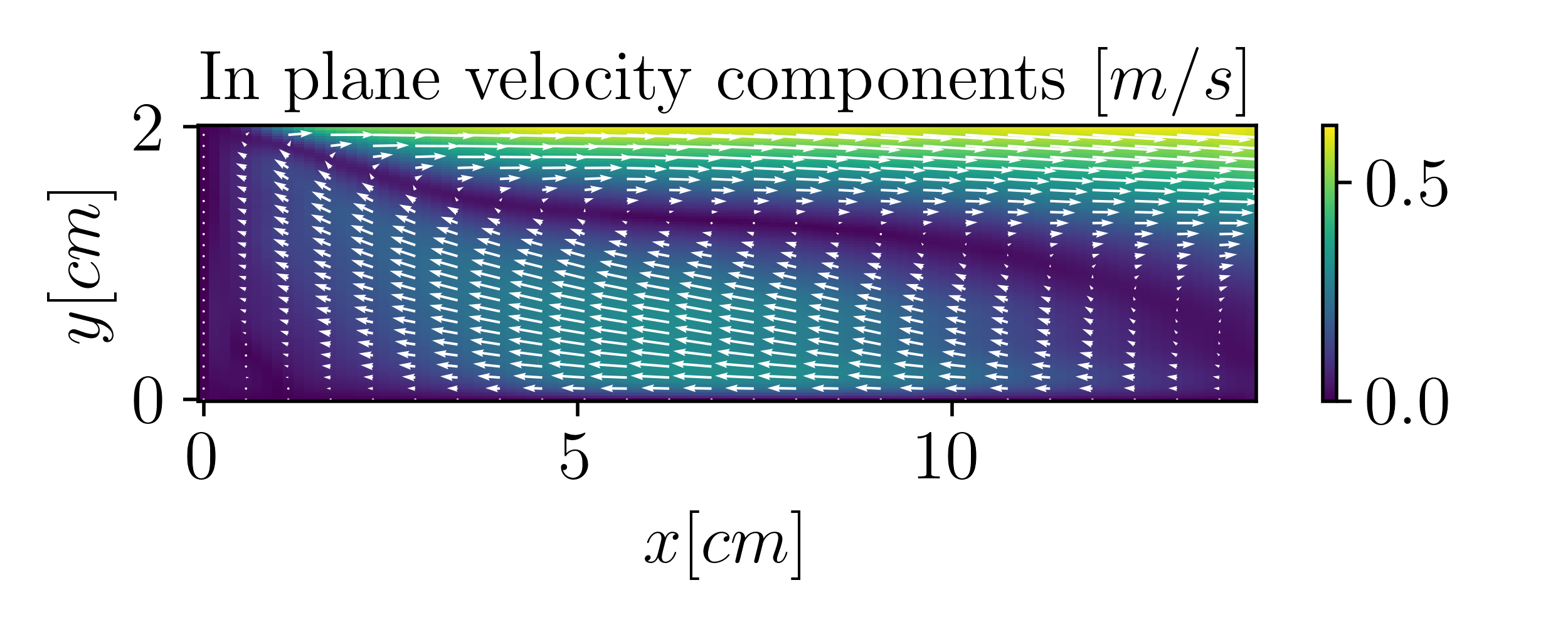}
    \caption{In plane velocity components $(u,v)$. DNS (left) and reconstructed (right)}
    \label{fig:uv}
\end{figure}

\begin{figure}
    \centering
    \includegraphics[width=0.48\textwidth]{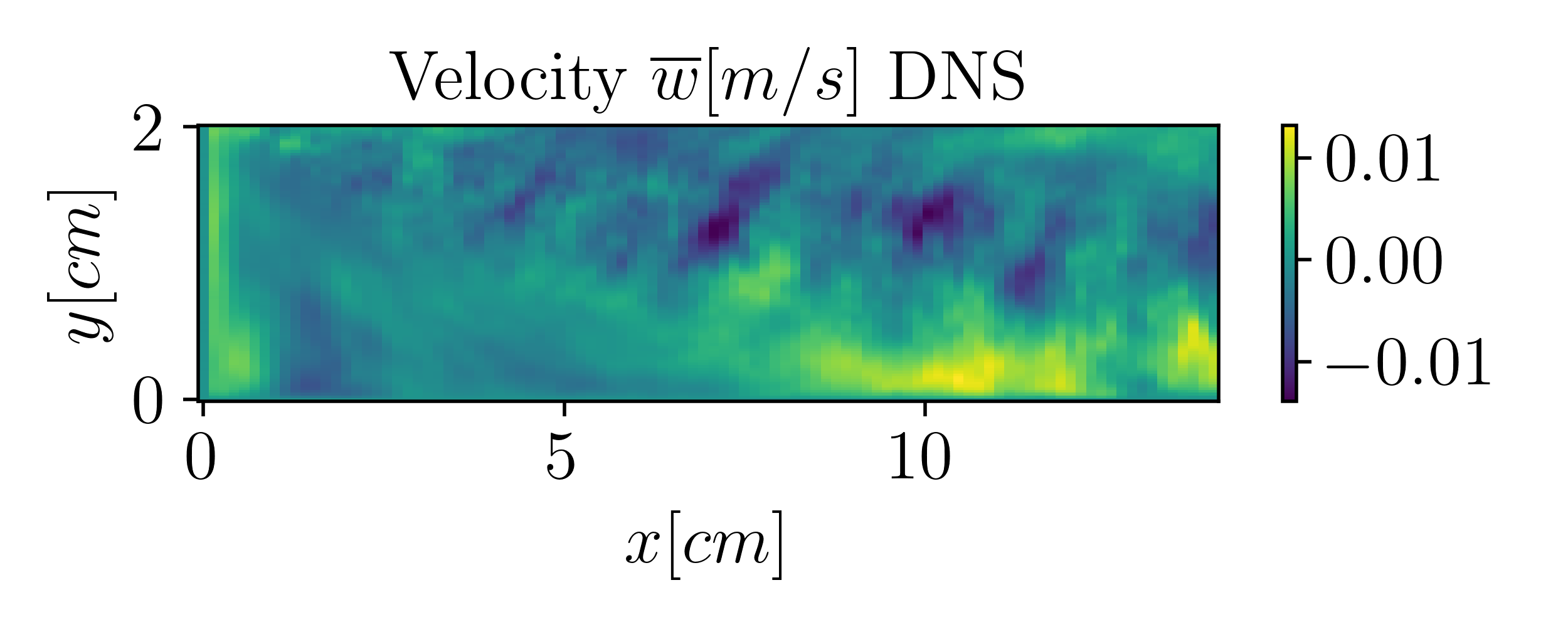}
    \includegraphics[width=0.48\textwidth]{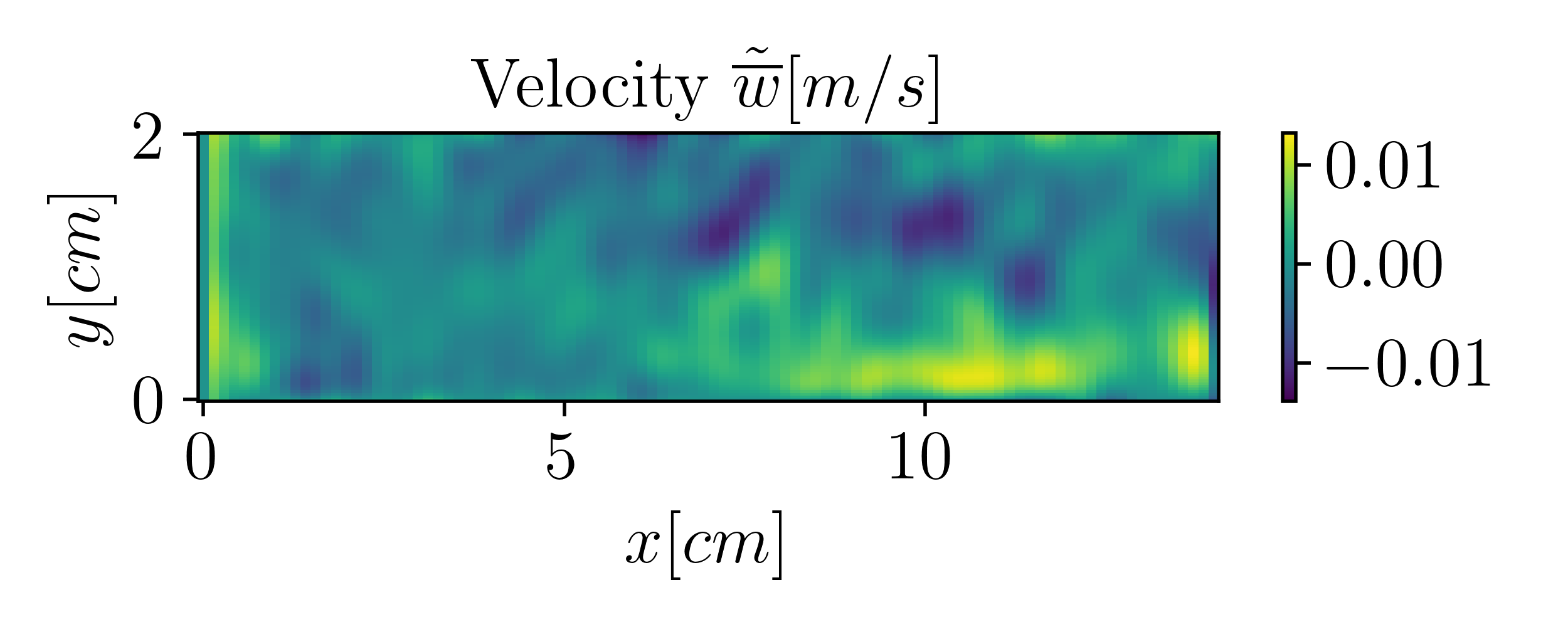}
    \caption{Span-wise velocity component $w$. DNS (left) and reconstructed (right)}
    \label{fig:w}
\end{figure}

\begin{figure}
    \centering
    \includegraphics[width=0.48\textwidth]{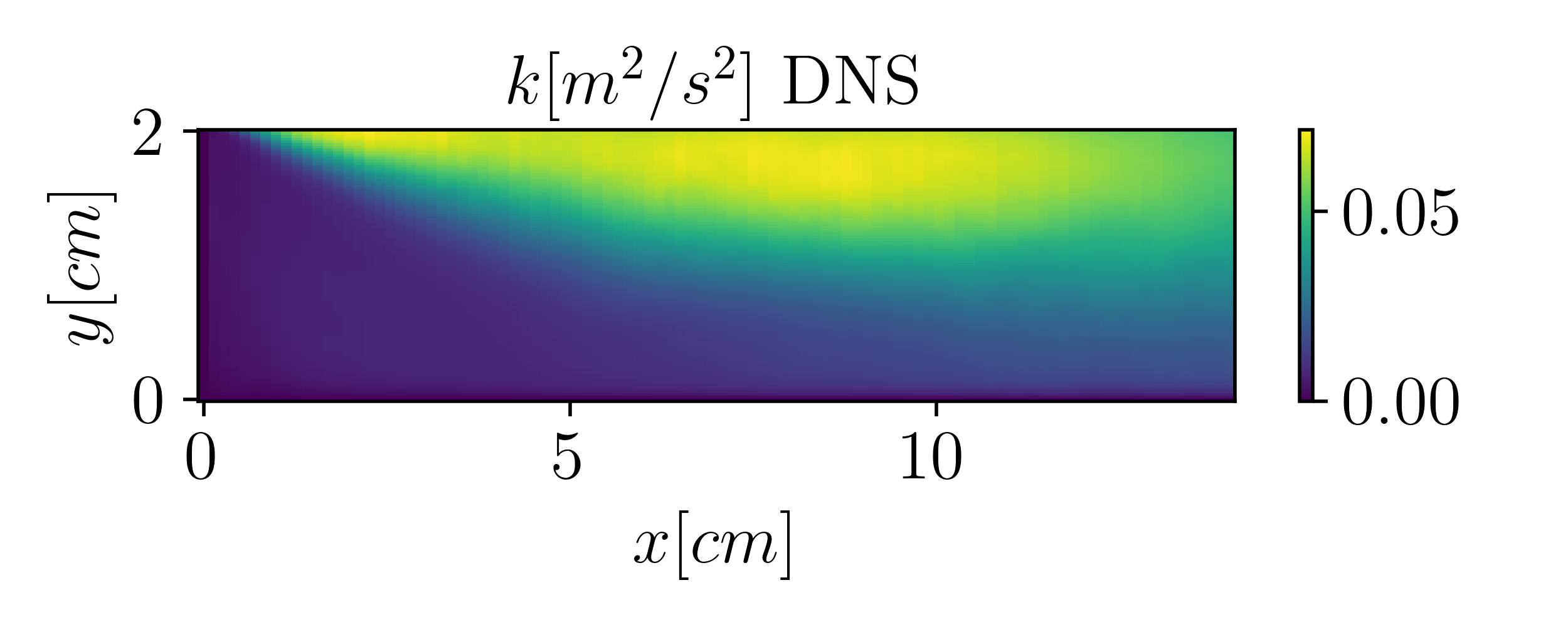}
    \includegraphics[width=0.48\textwidth]{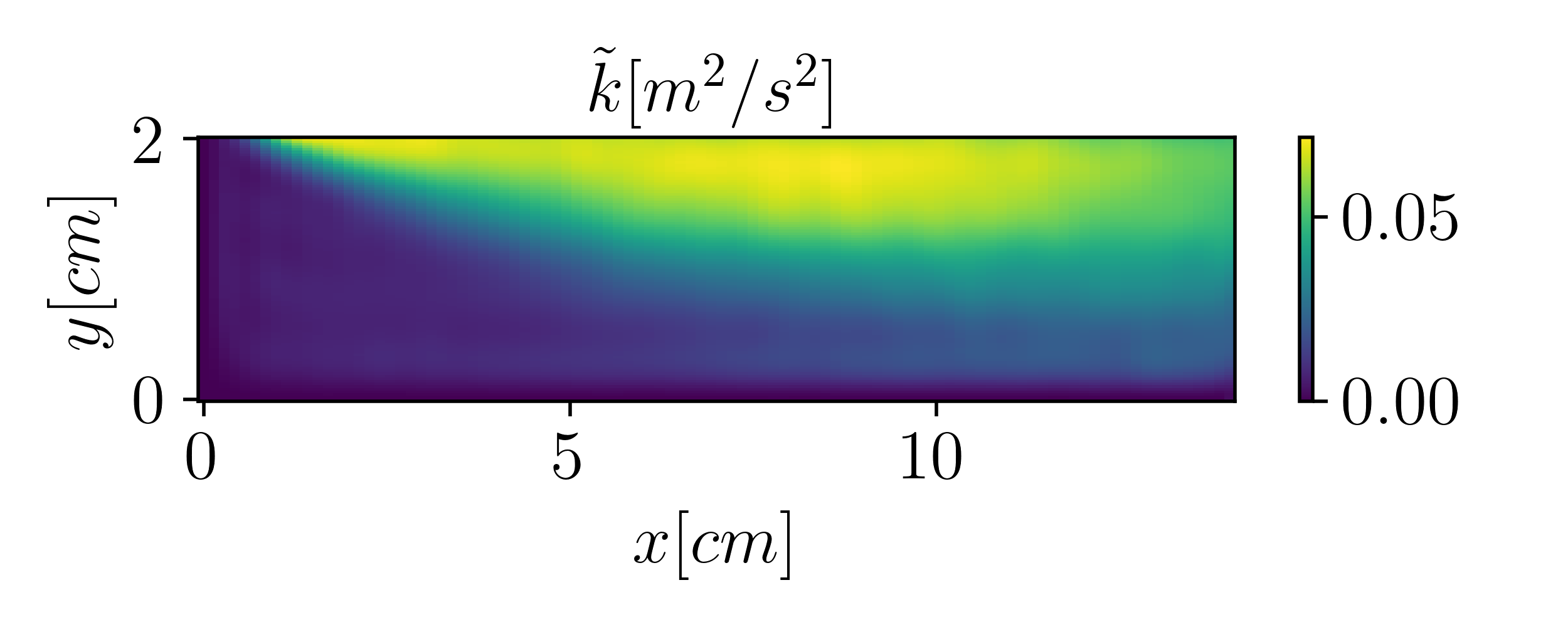}
    \caption{Turbulent kinetic energy $k$. DNS (left) and reconstructed (right)}
    \label{fig:k}
\end{figure}

\begin{figure}
    \centering
    \includegraphics[width=0.48\textwidth]{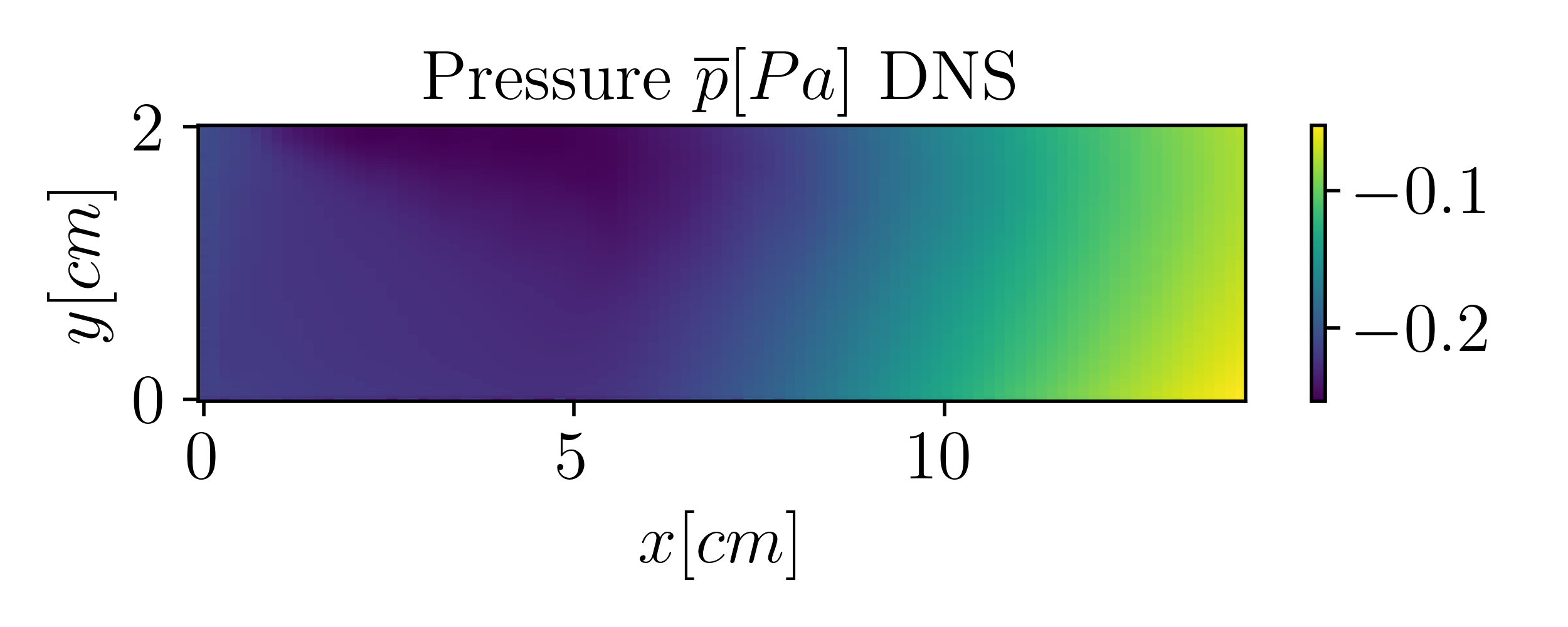}
    \includegraphics[width=0.48\textwidth]{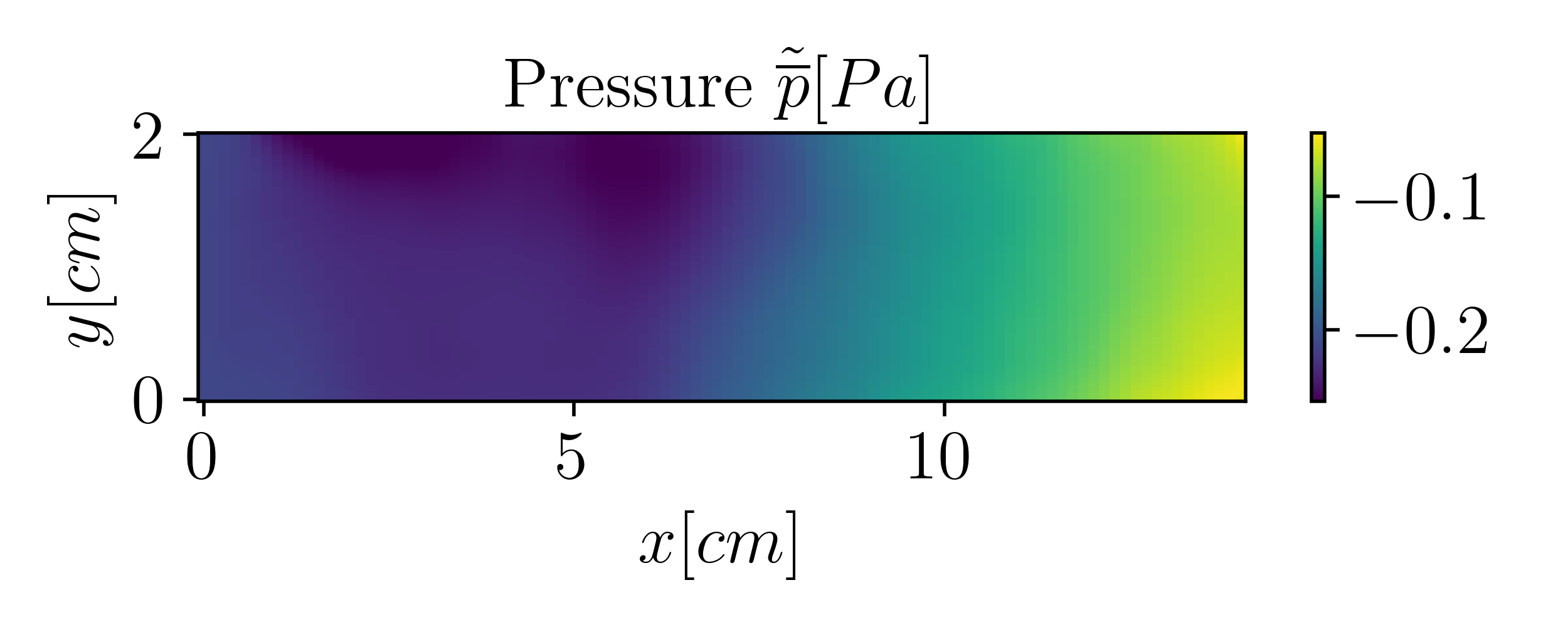}
    \caption{Pressure $p$. DNS (left) and reconstructed (right)}
    \label{fig:p}
\end{figure}

The result of the regression for the mean velocity are shown in Figure \ref{fig:uv} (right) and compared with the DNS data (left). These are indistinguishable. Figure \ref{fig:w} shows the same comparison for the velocity component $\overline{w}$, which particularly low values because of the bi-dimensionality of the flow. The reconstruction appears slightly smoothed but captures the main features. Finally, figures \ref{fig:k} and \ref{fig:p} shows the same comparison for the turbulent kinetic energy and pressure fields. Considering the moderate seeding density and the realistic noise levels, these results are considered largely satisfactory.

\begin{table}[h!]
    \centering
    \resizebox{0.95\textwidth}{!}{
    \begin{tabular}{|c|c|c|c|c|c|c|c|c|c|c|}
    \hline
         & $u$&$v$&$w$&$\overline{u'u'}$&$\overline{u'v'}$&$\overline{u'w'}$&$\overline{v'v'}$&$\overline{v'w'}$ &$\overline{w'w'}$&$p$\\\hline
      $\varepsilon$   &0.026&0.054&0.144&0.036&0.019&0.314&0.019&0.172&0.063&0.038\\\hline
      $E$   &0.571&0.156&0.206&0.111&0.021&0.013&0.033&0.028&0.128&0.668\\\hline
    \end{tabular}
    }
    \caption{Table of errors.}
    \label{tab:Errors}
\end{table}

The resulting in-plane velocities are shown in \figref{fig:uv}. Instead, the pressure reconstruction is displayed in \figref{fig:p}. The results are in agreement with the DNS flow field. Even though, they are compared over a uniform grid of size $100\times100$ in one plane resulting in $10000$ points. This value exceeds the original PTV data, sampled over the whole volume, by almost $2000$ points.

\section{Conclusion and future perspectives}

We proposed the first RANS extension of the RBF meshless integration framework for computing pressure fields from image velocimetry. This allows extending the meshless pressure computation to the mean pressure fields in turbulent flows. The method requires the statistical convergence of the velocimetry measurements and extends the work presented in \cite{Sperotto2022} by including the (constrained) regression of the Reynolds stresses. 

The meshless RANS integration was tested on a synthetic dataset constructed from DNS data, and no simplifying assumptions on the Reynolds stresses were considered (i.e. the turbulence is non-homogeneous and non-isotropic). The results of the pressure computation were excellent. Ongoing activities are focused on applying the method to PIV experimental data.

\bibliography{Sperotto_et_al_LX_2022}
\bibliographystyle{apacite}


\end{document}